# Epitaxial Growth of a Single-Crystal Hybridized Boron Nitride and Graphene layer on a Wide-Band Gap Semiconductor

Ha-Chul Shin,[†,‡] Yamujin Jang, [§,‡] Tae-Hoon Kim, [§,‡] Jun-Hae Lee,[†] Dong-Hwa Oh, [†] Sung Joon Ahn,[†] Jae Hyun Lee, [‖] Youngkwon Moon,[†] Ji-Hoon Park,[†] Sung Jong Yoo, [⊥] Chong-Yun Park,[†] Dongmok Whang,[*, §, ‖] Cheol-Woong Yang,[*, §] and Joung Real Ahn[*,†, ‖]

[†]Department of Physics, Sungkyunkwan University, Suwon 440-746, Republic of Korea,

[§]School of Advanced Materials Science and Engineering, Sungkyunkwan University, Suwon 440-746, Republic of Korea,

[‖]SAINT, Sungkyunkwan University, Suwon 440-746, Republic of Korea

[⊥]Fuel Cell Research Center, Korea Institute of Science and Technology (KIST), Seoul 136-791, Republic of Korea

* To whom correspondence should be addressed. E-mail: cwyang@skku.edu(C–W Yang), dwhang@skku.edu(D Whang), jrahn@skku.edu(JR Ahn)

‡These authors contributed equally.

**ABSTRACT**

Vertical and lateral heterogeneous structures of two-dimensional (2D) materials have paved the way for pioneering studies on the physics and applications of 2D materials. A hybridized hexagonal boron nitride (h-BN) and graphene lateral structure, a heterogeneous 2D structure, has been fabricated on single-crystal metals or metal foils by chemical vapor deposition (CVD). However, once fabricated on metals, the h-BN/graphene lateral structures require an additional transfer process for device applications, as reported for CVD graphene grown on metal foils. Here, we demonstrate that a single-crystal h-BN/graphene lateral structure can be epitaxially grown on a wide-gap semiconductor, SiC(0001). First, a single-crystal h-BN layer with the same orientation as bulk SiC was grown on a Si-terminated SiC substrate at 850 ℃ using borazine molecules. Second, when heated above 1150 ℃ in vacuum, the h-BN layer was partially removed and, subsequently, replaced with graphene domains. Interestingly, these graphene domains possess the same orientation as the h-BN layer, resulting in a single-crystal h-BN/graphene lateral structure on a whole sample area. For temperatures above 1600 ℃, the single-crystal h-BN layer was completely replaced by the single-crystal graphene layer. The crystalline structure, electronic band structure, and atomic structure of the h-BN/graphene lateral structure were studied by using low energy electron diffraction, angle-resolved photoemission spectroscopy, and scanning tunneling microscopy, respectively. The h-BN/graphene lateral structure fabricated on a wide-gap semiconductor substrate can be directly applied to devices without a further transfer process, as reported for epitaxial graphene on a SiC substrate.

**INTRODUCTION**

The superior physical properties of two-dimensional (2D) materials such as graphene, boron nitride (BN), and molybdenum disulfide ($MoS_2$) have significantly contributed to the innovation and development of electronic, photonic, and mechanical devices.[1,2] Recently, to overcome the limitations of homogeneous 2D materials, vertical and lateral heterogeneous 2D materials have also been studied. A graphene/h-BN vertical heterogeneous 2D material was first reported in ref 3. The graphene/h-BN vertical heterogeneous structure, when used instead of a graphene/$SiO_2$ structure, was shown to dramatically enhance the mobility of graphene. This large mobility enhancement was attributed to the presence of a large optical phonon energy, an atomically flat surface, and a homogenous charge distribution in the h-BN layer.[3] The mobility in a $MoS_2$ layer was also found to be higher in a $MoS_2$/h-BN vertical heterogeneous structure compared to that of a $MoS_2$/$SiO_2$ structure.[4,5] Graphene/$MoS_2$/graphene and graphene/h-BN/graphene vertical hybrid structures have been studied to develop vertical tunneling field effect transistors (FETs).[4] The hexagonal BN/graphene (h-BNC) lateral structure, a lateral heterogeneous 2D material, has also been actively investigated. Thus far, in the h-BNC structure, band gap opening,[6] an insulator-metal transition,[7] tunable electronics,[8] and a FET[9] have been experimentally reported. Theoretical proposals based on the h-BNC structure include antiferromagnetism,[10] unique thermal transport phenomena,[11] and a lateral tunneling FET.[12]

The h-BNC layer was first fabricated on Cu foil using a thermal catalytic chemical vapor deposition (CVD) process, where methane ($CH_4$) and ammonia borane ($NH_3$-$BH_3$) molecules were used as precursors for carbon and boron nitride.[6] Afterward, the domain sizes of graphene and h-BN were successfully controlled.[6,7] Lateral interface structures between graphene and h-BN domains were studied on metal crystals, specifically Ru(0001)[13] and Rh(111)[14]. Recently, h-BNC was fabricated using a high-temperature topochemical conversion technique that converts graphene to h-BNC and h-BN layers on a high-resistance intrinsic silicon substrate, where the graphene was grown on Cu foil and, subsequently, transferred onto the silicon substrate.[9] For homogeneous graphene, two representative types of graphene, CVD graphene on metal foils and epitaxial graphene on a SiC substrate, have been applied to graphene devices.[15-16] CVD graphene on a metal foil can be grown up to a very large scale, but is polycrystalline[17,18] and must be transferred onto an insulating substrate, such as a $SiO_2$/Si substrate, for device applications.[15,19] Epitaxial graphene grown on a wide band-gap semiconductor substrate made of SiC has a size limited to that of the SiC substrate, but is single-crystal and does not require further transfer processes for device applications.[20-16,21] The h-BNC layer, as described above, has been synthesized on metal substrates, in which case an additional process that transfers the h-BNC layer onto an insulating substrate is required for device applications. Therefore, the challenge remains to achieve a epitaxial growth of single-crystal h-BNC directly located on a wide-band gap semiconductor substrate.

Here we demonstrate that a single-crystal h-BNC layer can be epitaxially grown on a wide-band gap semiconductor substrate. In this demonstration, a SiC(0001) was used as the wide-band gap semiconductor substrate.[21] First, a single-crystal h-BN layer with the same orientation as the bulk SiC, denoted by R0°, was grown on a Si-terminated SiC substrate with a superstructure of $(\sqrt{3} \times \sqrt{3})R30°$ using a thermal catalytic CVD, where borazine ($B_3N_3H_6$) molecules were used as a precursor. When the h-BN covered SiC substrate was thermally heated to a temperature above 1150 °C, the h-BN layer was partially decomposed and, sequentially, a graphene domain replaced the h-BN region. Interestingly, the graphene domain had the same orientation as the h-BN layer, resulting in a single-crystal h-BNC layer with orientation R0°; a typical epitaxial graphene on a SiC(0001) substrate has a rotational angle of 30°, denoted by R30°, with respect to the bulk SiC.[22-23] Finally, when the single-crystal h-BNC layer was heated above 1600 °C, single-crystal graphene with orientation R0° completely replaced the single-crystal h-BN with R0°. The crystalline structure of the h-BNC layer was confirmed by using low energy electron diffraction (LEED) with an electron beam size of approximately 1 mm. The characteristic electronic band structures of h-BN, h-BNC, and graphene layers were measured by using *k*-resolved photoemission spectroscopy (PES). The spatial distributions of the h-BN and graphene domains in the h-BNC layer were observed by using scanning tunneling microscopy (STM), transmission electron microscopy (TEM), and scanning electron microscopy (SEM). h-BNC devices were directly fabricated on SiC without transfer process to demonstrate a device application of the h-BNC layer and used to measure the resistance of h-BNC channels.

**RESULTS AND DISCUSSION**

Figure 1 shows the LEED patterns of h-BN, h-BNC, and graphene layers with an electron beam energy of 60 eV. The LEED pattern presented in Figure 1b,c clearly shows a (5 × 5) moire superstructure of a single-crystal h-BN layer on a SiC(0001) substrate, where the h-BN layer and bulk SiC have the same orientation. The moiré superstructure shows that the lattice constant of the h-BN layer is approximately 2.56 Å, approximately 2% larger than the in-plane lattice constant of bulk h-BN, 2.51 Å.[24,25] Thus, the moiré superstructure suggests that the h-BN layer is under tensile stress on the SiC substrate. When the single-crystal h-BN layer was heated above 1150 °C, the spot positions in the LEED pattern gradually moved from that of 2% stretched h-BN to graphene, and subsequently to a pattern corresponding to unstrained h-BN; the (5 × 5) LEED pattern gradually disappeared and, subsequently, a moiré LEED pattern with R0° around the LEED spots of graphene emerged (see Figure 1d-f and Figure S1 in Supporting Information). The graphene with orientation R0° with respect to bulk SiC is very interesting compared to typical epitaxial graphene grown on a SiC substrate. When graphene is grown on a SiC substrate only by using thermal heating, it has R30° with respect to bulk SiC.[22] Hence, the single-crystal graphene with R0° must be related to the single-crystal h-BN layer with R0°. When a graphene domain is grown in a h-BN layer, the graphene domain may energetically prefer to keep the same orientation as the h-BN layer; otherwise, defective domain boundaries can be created.[17,26] The LEED experiments suggest that a single-crystal h-BNC layer can be fabricated at temperatures ranging from 1150 to 1600 °C. On the other hand, the moiré LEED pattern around the LEED spot corresponding to R0° graphene cannot be understood when the graphene is directly located on a bulk-terminated structure of SiC; in this case a moiré LEED pattern should be observed around the LEED spot of bulk SiC. The moiré LEED pattern suggests that an interface structure exists between the graphene and the bulk SiC. When LEED was performed with a 100 eV electron beam (see Figure S2 in Supporting Information), another moiré pattern around the LEED spot of the bulk SiC peak was observed. The moiré pattern is $(6\sqrt{3} \times 6\sqrt{3})R30°$,[22,27] which corresponds to a zeroth layer (or buffer layer) with the same structure as the graphene with R30°. The moiré pattern suggests that graphene with R30°, which is chemically bonded to the bulk SiC, exists as an interface layer between the graphene with R0° and the bulk SiC (Figure S2c in Supporting Information). The existence of this interface layer was confirmed by the following hydrogen interaction. After hydrogen intercalation, the moiré patterns around both the LEED spots of the bulk SiC and the graphene with R0° disappeared, and the LEED pattern of graphene with R30° was observed (see Figure S2d-f in Supporting Information).[27] The hydrogen intercalation suggests that the interface structure is a zeroth-layer graphene with R30°, which is chemically bonded to the bulk SiC.

As the $(5 \times 5)$ moire superstructure undergoes a phase transition into the graphene with R0°, changes in electronic band structure were observed by using *k*-resolved PES (Figure 2 and 3). Figure 2b shows the electronic band structure of the $(5 \times 5)$ moire superstructure in the $\overline{\Gamma} - \overline{K}$ direction, denoted by $k_x$, where the $\overline{K}$ points of stretched h-BN, pristine h-BN, and graphene are denoted by $\overline{K}'_{BN}$, $\overline{K}_{BN}$, and $\overline{K}_G$, respectively. Three dispersive energy bands were observed in the PES intensity maps shown in Figure 2b: the two dispersive energy bands denoted by the white curves correspond to the σ bands of h-BN, while the other energy band denoted by the yellow curve corresponds to the π band of h-BN.[25,28] The π and σ bands were slightly distorted from the intrinsic bands of h-BN. The difference between the π band maximum at the $\overline{K}'_{BN}$ point and the σ band maximum at the $\overline{\Gamma}$ point is -0.3 eV, while that of pristine h-BN calculated using density functional theory (DFT) is 1.3 eV.[28] The deviation of the electronic band structure of 2%-stretched h-BN from that of bare h-BN can originate from tensile stress induced by the stretching or interactions with an underlying SiC substrate. We thus performed first-principles calculations to understand the electronic band structures of bare and 2%-stretched h-BN (Figure S3 in Supporting Information). In the valence band, the difference between the π band maximum at the $\overline{K}_{BN}$ point and the σ band maximum at the $\overline{\Gamma}$ point of bare h-BN is nearly the same as that of 2%-stretched h-BN. This result suggests that the difference between the experimental and theoretical electronic band structures of 2%-stretched h-BN can originate from interactions with an underlying SiC substrate rather than tensile stress induced by the stretching. The *k*-resolved PES and LEED experiments clearly indicate that a single-crystal h-BN layer was directly grown on SiC without a buffer layer, which is supported by TEM experiments, as described below, while a h-BN film was previously reported to be grown on a BN buffer layer with a thickness of 20 nm on SiC using low pressure metal organic CVD.[35] When the single-crystal h-BN layer was heated above 1150 °C (Figure 2c), the π band of graphene (red curve) was observed.[29,23] Detailed changes in the electronic band structures near the $\overline{K}$ points along the $k_y$ direction were measured with increasing temperature (Figure 3). The π band of graphene became visible after the sample was heated to a temperature of 1150 °C, and sharpened at higher temperatures, while the intensity of the π band for h-BN gradually reduced and eventually disappeared. The increased intensity of the π band for graphene and the decreased intensity of the π band for h-BN with increasing temperature suggest that at higher temperatures, graphene domains are created and enlarged while the area of the h-BN layer is reduced, resulting in a h-BNC layer with different ratios of h-BN and graphene domains. However, it is hard to confirm band gap of graphene induced by quantum confinement using by ARPES, because ARPES usually measure in large area and the band broadening occur due to incoherent graphene domain shape and size. When the π band of h-BN disappeared at 1600 °C, the Dirac point of the π band for graphene was located at the binding energy of approximately 0.2 eV, where the Dirac point was determined from energy distribution curves (see Figure S4 in Supporting Information). The Dirac point energy indicates that the graphene with orientation R0° is *n*-type, where electrons were transferred to graphene from an underlying SiC structure. Compared to R0° graphene, typical epitaxial graphene with R30°, grown only by thermal heating, has a Dirac point energy of 0.4 eV.[23] The difference in Dirac points suggests that an interaction between an underlying structure and the R0° graphene may be different from the same interaction with R30° graphene. Furthermore, the binding energy of the π band of h-BN near the $\overline{K}$ point shifted from 4.06 to 3.1 eV as the temperature increased (Figure 3a-e). When heated to 1350 °C, the difference between the π band maximum at the $\overline{K}_{BN}$ point and the σ band maximum at the $\overline{\mathit{\Gamma}}$ point became similar to that of bare h-BN (Figure 2c). The change in the π band of h-BN suggests that strain, which is applied to h-BN and/or an underlying structure of h-BN, changes when a h-BNC layer forms at temperatures between 1150°C and 1600°C.

To understand the spatial distributions of the h-BN and graphene domains at different temperatures, STM images of the sample were acquired (Figure 4). Figure 4a shows a STM image of the $(5 \times 5)$ moiré superstructure for the h-BN layer that was grown at 850 °C using borazine molecules. When the h-BN layer was heated to temperatures above 1150 °C (Figure 4b), brighter regions (region I in Figure 4f,g), which are dominant and correspond to the h-BN domains, were randomly and partially removed; consequently, darker regions (region II in Figure 4f,g) were created. The darker region corresponding to the graphene domain in Figure 4g shows a typical $(\sqrt{3} \times \sqrt{3})R30°$ modulation near linear defects due to electron scattering at the graphene edges (see Figure S5 in Supporting Information).[30] A two-dimensional fast Fourier transformation (2D-FFT) of the STM image showing the $(\sqrt{3} \times \sqrt{3})R30°$ modulation is shown in Figure S5 in Supporting Information. In the 2D-FFT, the $(\sqrt{3} \times \sqrt{3})R30°$ modulation is clearly observed. Furthermore, the STM image in Figure 4g shows that the graphene domain is atomically connected with the brighter region corresponding to the h-BN domain. The darker regions were enlarged at higher temperatures (Figure 4b-d). When heated to 1600°C, the darker regions completely replaced the brighter regions. Figure 4e shows that the darker region, which was measured after heating to 1600 °C, corresponds to the moiré superstructure of graphene with R0°. The height difference between the h-BN and the graphene domains is approximately 0.23 nm. This is discussed in more details in below with TEM cross-section images. The relative frequencies were taken as heights in the STM images to determine the area ratio of the h-BN to the graphene domains, where the h-BN and graphene domains are located at zero and -0.23 nm, respectively (Figure 4h). The height frequencies clearly show that the total area of the graphene domains increases while that of the h-BN domains decreases as the temperature increases. The area ratios of the h-BN to graphene domains, which were determined from STM images, were compared with the intensity ratios of the π band of h-BN to that of graphene, which were determined from *k*-resolved PES intensity maps at the $\bar{K}$ points (Figure 4i). The ratios indicate that the *k*-resolved PES experiments are consistent with the STM experiments. As shown in STM images, when a h-BN layer was partially replaced by graphene domains, the overall structure of the graphene domains resembles a 2D graphene random network that is composed of graphene ribbons, where graphene domains are connected with each other through graphene ribbons. As reported previously, when the width of a graphene ribbon is small enough, it becomes semiconducting with a finite energy gap because of quantum confinement.[37] At the initial stage of graphene growth, the semiconducting graphene domains could be predominant, but as the total area of graphene domains increases, the metallic graphene domains could become more dominant. On the h-BNC layer, graphene ribbons with different widths coexist, as observed in STM images. Hence, in ARPES experiments, the electronic band structures of graphene domains with different widths are overlapped so that the energy gap opening could not be clearly resolved.

To understand the underlying structures of h-BN and graphene domains, we performed cross-section TEM experiments (Figure 5 and Figure S6 in in Supporting Information). The height difference between h-BN and graphene domains is 0.23 nm, which is consistent with STM experiments. We draw line profiles across the layers to measure the exact positions of the layers, as shown in Figure 5c. The blue and red line profiles were obtained at h-BN (region I) and graphene (region II) domains, respectively. As shown in Figure 5, the bulk SiC of a h-BN domain are terminated at a different position from that of a graphene domain. The height difference between h-BN and graphene domains thus originates mainly from the difference positions of the topmost SiC layers. Furthermore, interactions of the h-BN layer with the underlying SiC structure is different from those of the graphene layer. As a result, the interlayer distance between the h-BN and the topmost SiC layers is 0.37 nm, while the interlayer distance is 0.40 nm for the graphene layer. The different interlayer distances also contribute to the height difference between h-BN and graphene domains. On the other hand, if both domains were h-BN or graphene, the interlayer distances should be the same. Therefore, the different interlayer distances support that the domains have different origins, as described above. In addition, TEM images clearly show that the h-BN and graphene domains are single-layer and atomically connected to each other. We got the same TEM images for a different sample, as shown in Figure S6 in Supporting Information.

We also obtained SEM images of the h-BNC layer, where h-BN and graphene domains can be imaged differently because of different work functions (Figure S7b in Supporting Information). In a previous report on SEM experiments of a h-BNC layer on Ru(0001), dark and bright images were assigned to graphene and h-BN domains, respectively.[13,14] Hence, the dark and bright images in the SEM image in Figure S7b in Supporting Information could be assigned to graphene and h-BN domains, respectively. The shape of the bright image of the SEM image is very similar to the bright image corresponding to the h-BN of the STM image (Figure S7a in Supporting Information), where we used samples that were grown under the same condition for SEM and STM experiments. Furthermore, the step edges were also observed in the SEM image (Figure S7b in Supporting Information), as indicated by the black arrows, and have similar widths to those observed in the STM image (Figure S7a in Supporting Information). Therefore, both cross-section TEM and SEM images support the existence of the h-BNC layer.

Based on the LEED, *k*-resolved PES, STM and TEM experiments, the schematic growth mechanism of h-BNC is illustrated in Figure 6. The reactive Si adatoms on the Si-terminated SiC substrate with a $(\sqrt{3} \times \sqrt{3})$R30° superstructure may act as a catalyst for the decomposition of borazine molecules into boron and carbon atoms. The catalytic decomposition of borazine results in the formation of a single-crystal h-BN layer with orientation R0°. When a Si-terminated SiC substrate with more Si atoms, a $(3 \times 3)$ superstrucure, was used as a catalyst, a h-BN layer was not grown. Furthermore, when a C-terminated SiC substrate with a $(6\sqrt{3} \times 6\sqrt{3})$R30° superstructure was chosen, the h-BN layer was also not observed. These results suggest that the Si-terminated SiC substrate with a $(\sqrt{3} \times \sqrt{3})$R30° superstructure is an optimized catalytic surface for the growth of h-BN. When the h-BN layer was heated to higher temperatures, the h-BN layer decomposed and a graphene domain with R0° replaced the decomposed h-BN region. Graphene domains with R30° are grown on 6H-SiC(0001) when a h-BN layer is not used. In our experiments, when the h-BN layer was fully grown, we observed only graphene domains with R0° in STM images and LEED patterns. Therefore, the results suggest that lateral interactions between graphene and h-BN layers are much stronger than vertical interactions between a graphene layer and bulk SiC. The growth mechanism of graphene with R0° is different from that of typical epitaxial graphene with R30° grown on SiC using only thermal silicon sublimation. The R0° graphene domains randomly grow on SiC terraces. Contrary to the R0° graphene, typical epitaxial graphene with R30° grows from SiC step edges where Si atoms are highly sublimated, resulting in the coexistence of graphene domains with different layers.[38-42] For graphene with orientation R0°, even under heating to a temperature of 1600 °C in ultra high vacuum, only the typical electronic band structure of single-layer graphene was observed in *k*-resolved PES experiments. The uniform growth of single-layer graphene with R0° at a high temperature, compared to epitaxial graphene with R30°, can be explained in terms of the suppressed sublimation of Si atoms by the h-BN layer. The growth mechanism may be similar to that for graphene fabricated in argon or disilene environments.[32,33] The height difference between the h-BN and graphene domains is 0.23 nm, and the two domains are atomically connected to each other with the same orientation. Therefore, the h-BNC layer can be schematically drawn as shown in Figure 6c,d. The interface structure, called a zeroth-layer graphene with R30°, has a superstructure of $(6\sqrt{3} \times 6\sqrt{3})R30°$. Therefore, the existence of the interface structure can be determined from LEED patterns. When a h-BN layer was partially replaced by graphene domains with R0°, the $(6\sqrt{3} \times 6\sqrt{3})R30°$ LEED pattern was not observed on the h-BNC layers. The $(6\sqrt{3} \times 6\sqrt{3})R30°$ LEED pattern was observed when the h-BN layer was fully replaced by graphene domains with R0° at higher temperatures. The results suggest that the interface structure formed at higher temperature after graphene domains with R0° was grown. Therefore, in the TEM images of the h-BNC layers, the interface structure with a superstructure of $(6\sqrt{3} \times 6\sqrt{3})R30°$ was not observed, which is consistent with LEED patterns.

On the other hand, single-crystal graphene was recently reported to be grown on a Ge(110) film using CVD, where the Ge(110) film was grown on a Si wafer.[43] In comparison to the CVD-grown single-crystal graphene, we suggest in this report that single-crystal h-BNC (or graphene) can be also grown on a SiC substrate. The maximum size of a commercial Si wafer is much larger than that of a commercial SiC wafer. It is thus better to fabricate large-scale single-crystal graphene based on a Ge film grown on a Si wafer. However, for device applications, graphene grown on a Ge film should be transferred onto an insulator substrate such as a SiO2/Si wafer, while h-BNC (or graphene) grown on a SiC substrate does not require a transfer process for device applications.

To demonstrate the direct fabrication of a h-BNC device on SiC without transfer process, we directly fabricated a h-BNC channel on SiC without transfer process and, subsequently, made a source and a drain on the h-BNC channel (Figure 7). We measured the resistance of the h-BNC channel using the device as a function of a heating temperature, where the schematic structure of the h-BNC device is drawn in Figure 7a and the total area of graphene domains increase as the heating temperature is raised, as described above. The resistance measurements were performed at 100 K to freeze the dopants of SiC, where a bare SiC sample was testified to confirm the freezing of the dopants at 100 K and found to be completely insulating. Figure 7b shows the I-V curves of the h-BNC channels. The results show that the resistance increases as the total area of the graphene domains increases, which is consistent with the previous report of h-BNC on $SiO_2$.[6]

We also fabricated FETs based on the h-BNC channels, as shown in Figure S8 in Supporting Information. We fabricated FETs using a top gate, where $Al_2O_3$ was used as a gate oxide. The h-BNC layer grown on a SiC substrate resembles a 2D random network of graphene and h-BN domains. Such a h-BNC layer was observed when a h-BNC layer was grown by CVD.[6] The h-BNC layer was fabricated on a Cu foil using methane ($CH_4$) and ammonia borane ($NH_3$-$BH_3$). The CVD-grown h-BNC layer was transferred onto a $SiO_2$/Si substrate for device applications. FETs based on the CVD-grown h-BNC layer were fabricated using a back gate. The mobilities of the FETs based on the CVD-grown h-BNC layer were between $5$ and $20\,cm^2\,V^{-1}s^{-1}$, while the on/off ratios of the FETs were not clearly mentioned. The on/off ratios of the FETs, based on the back gate voltage-drain current curve, is roughly 1.5. The mobilities of the FETs based on the h-BNC layer are much smaller than the mobility of the FETs based on graphene because of electron scattering at the boundary between h-BN and graphene domains.[6] The mobilities of the FETs based on the h-BNC layers grown on SiC substrates in this report was between $0.3$ and $8.9\,cm^2\,V^{-1}s^{-1}$ while the on/off ratios of the FETs were approximately 1.2. The mobilities and on/off ratios of the FETs based on the h-BNC layer grown on a SiC substrate are comparable with those of the FETs base on the CVD-grown h-BNC layer. The fabrication of the top gate oxide of the FET based on the h-BNC layer grown on a SiC substrate can degrade the quality of the h-BNC channel, while the FETs based the CVD-grown h-BNC layer is based on a back gate. Therefore, it is reasonable that the mobilities of the FETs of the h-BNC layers grown on SiC substrates are slightly smaller than those of the CVD-grown h-BNC layer.

**CONCLUSION**

We demonstrate that a single-crystal hybridized h-BN/graphene layer can be grown epitaxially on a wide gap semiconductor substrate, SiC(0001). A single-crystal h-BN layer with orientation R0° was epitaxially grown at 850 ℃ using borazine molecules. Single-crystal hybridized h-BN/graphene layers with R0° were fabricated when the h-BN layer was heated at temperatures ranging from 1150 to 1600 ℃, where the graphene domain gradually replaced the h-BN layer while maintaining the same orientation. The single-crystal graphene layer with R0° completely replaced the h-BN layer when heated at 1600 ℃. Hence, the area ratio of the h-BN to graphene domains can be controlled by temperature. The height difference between the h-BN and graphene domains is approximately 0.23 nm, and the h-BN and graphene domains are atomically connected to each other. Figure 6a shows that I-V curves for h-BNC on SiC(0001) depend on heating temperature which is related ratio between h-BN and graphene domains. The resistance decrease when heating temperature (graphene domain) increases (Figure 6b).[9] The epitaxial, single-crystal hybridized h-BN/graphene layer on a wide-gap semiconductor substrate can facilitate device applications for the hybridized structure without a transfer process, as demonstrated for epitaxial graphene on a SiC substrate.

**METHODS**

**Sample Preparation**. A 6H- or 4H-SiC(0001) substrate with a size of $10 \times 3$ mm$^2$ was used for experiments because the size is available to our UHV system. (Figure S9 in Supporting Information). The SiC sample was hydrogen-etched and subsequently loaded into a UHV chamber. A $(3 \times 3)$ superstructure was observed after heating the sample to $850\ °C$ with Si flux. Sequent thermal heating at $900\ °C$ without Si flux resulted in a Si-terminated SiC(0001) substrate with a $(\sqrt{3} \times \sqrt{3})$R30° superstructure, where Si atoms were thermally evaporated under the heating process, as shown in Figure 1a. A Single-crystal h-BN layer was grown by exposing the Si-terminated SiC substrate with a $(\sqrt{3} \times \sqrt{3})$R30° superstructure to borazine at a pressure of $10^{-5}$ torr and a temperature of $850\ °C$ for 5 minutes. After h-BN synthesis, h-BNCs and graphene with orientation R0° were fabricated by converting from h-BN to graphene domains after annealing at $1150\ °C$, $1250\ °C$, $1350\ °C$, $1450\ °C$ and $1600\ °C$ for 5 minutes. LEED, ARPES and STM measurements were performed after reaching each of these temperatures.

**ARPES experiments.** The ARPES spectra were measured with a commercial angle-resolved photoelectron spectrometer (R3000, VG-Scienta) using monochromated He-II radiation (hυ = 40.8 eV, VG-Scienta) at RT. The base pressure was less than $5.0 \times 10^{-11}$ Torr, and the overall energy and angular resolutions were $70$ meV and $0.1°$, respectively.

**STM experiment**s. The STM images were obtained using VT-STM (Omicron) at RT in vacuum with a pressure less than $2.0 \times 10^{-11}$ Torr.

**TEM experiments.** The preparation of TEM samples and the TEM measurements are as in the following. A cross-section sample for TEM analysis was fabricated using a triple-beam instrument (SIINT SMI3050TB) and the lift-out approach. The instrument combines a Ga ion beam, a SEM for process monitoring, and an Ar ion beam to remove the layers damaged by focused ion beam (FIB). To prevent the damage of the h-BNC layer during Ga ion milling, a protective layer was deposited on the surface and low-kV Ga and Ar ion milling have been performed. Scanning TEM (STEM) imaging was carried out using an aberration corrected JEM-ARM200F operated at 80kV. TEM Images were collected using a convergence semi-angle of 21 mrad and bright-field (BF) detector with a collection semi-angle of 0-17 mrad.

**SEM experiments.** SEM imaging was carried out using JSM-7500F operated at 0.8 kV with gentle beam (GB) mode decelerates incident electrons just before they hit the specimen to reduce the incident-electron penetration and the charging in the specimen. The GB mode provides high-resolution images whose quality is as high as those of higher accelerating voltages, even at low accelerating voltage down to 100 V without damaging the specimen surface.

**Device fabrications.** The fabrication process of the h-BNC device is as in the following. First, a Au film of 50 nm was deposited using a thermal evaporator. Second, a Au pattern was prepared by etching process using photolithography. The Au patterns was used as a source and a drain.[43] $Al_2O_3$ gate oxides with a thickness of 50 nm were grown using atomic layer deposition at 200 °C. Cr/Au top gate electrodes were patterned using photolithography, where the thicknesses of the Cr and Au films were 5 and 50 nm, respectively. The device measurements were performed using the Keithley 4200-SCS semiconductor characterization system at 100K.

**First principle calculations.** We carried out first-principles calculations based on density functional theory using vasp code.[44-47] The potentials for electron-electron and electron-ion interactions were described projected augment method[48,49] and generalized gradient approximation of Perdew, Burke and Ernzerhof version.[50,51] The electronic wave function was linear combination of plane waves, which the kinetic energy cutoff of 400 eV. The unit cell contains a B atom and a N atoms and 8 Å vacuum. For electronic density, we used 9 ×9 Monkhorst and Pack mesh[51] in surface Brillouin zone.

**ACKNOWLEDGEMENTS**

This study was supported by a grant from the National Research Foundation of Korea (NRF), funded by the Korean government (MEST) (No. 2012R1A1A2041241). D.W. acknowledges the support by Basic Science Research Program through NRF (2009-0083540). C.Y. acknowledges the support by NRF (No. 2011-0019984).

**SUPPORTING INFORMATION AVAILABLE**

Line profiles of LEED patterns (S1), the LEED pattern of graphene with R0° on SiC at an electron beam energy of 100 eV before and after hydrogen intercalation (S2), electronic band structures of bare h-BN and 2%streched h-BN(S3), the energy distribution curves (EDCs) of graphene along the $k_y$ direction after heating at 1600℃ (S4), a 2D FFT image of Figure 4g (S5), TEM cross-section images of h-BNC on SiC(0001) (S6) and STM and SEM images of h-BNC on SiC after heating at 1450 ℃ (S7). The source-drain currents as a function of voltage applied to the top gate for h-BNC device grown on SiC (S8). The optical images of samples grown on SiC dice (S9). This material is available free of charge via the Internet at http://pubs.acs.org.

**FIGURES**

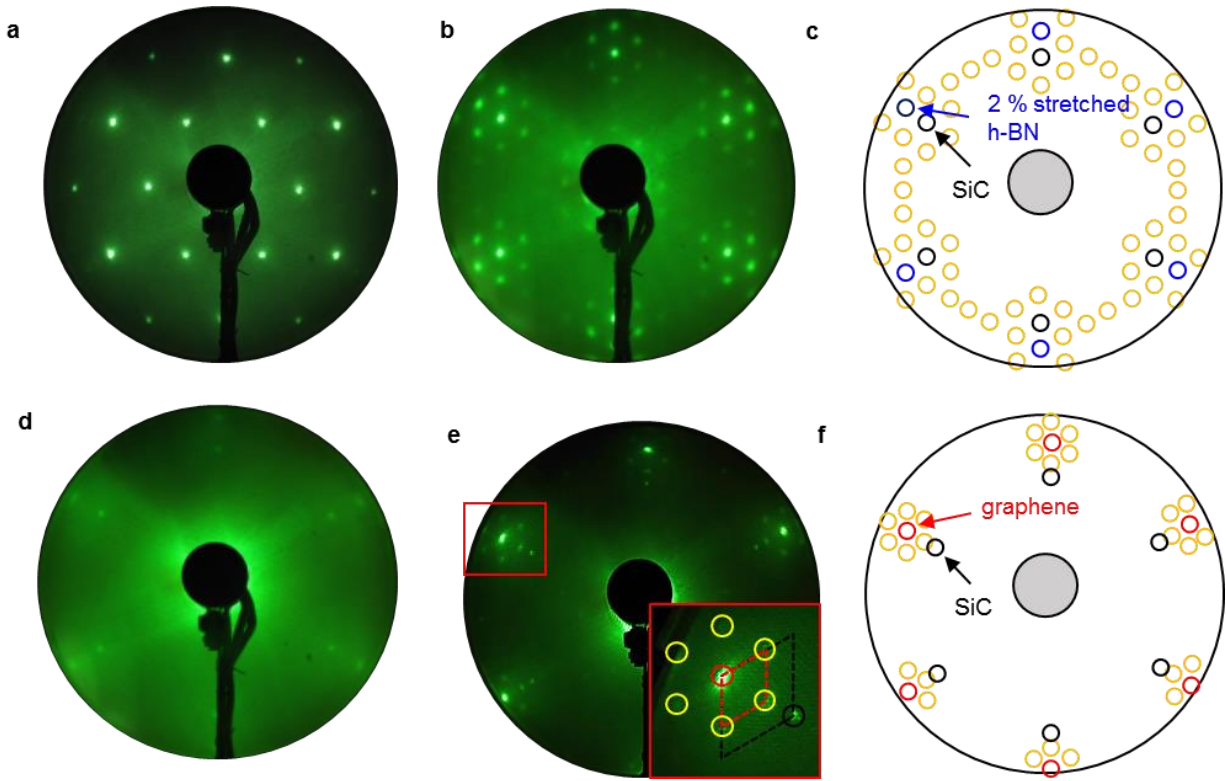

**Figure 1. LEED patterns of h-BN, h-BNC, and graphene.** (a) $(\sqrt{3} \times \sqrt{3})R30°$ LEED pattern of a Si-terminated SiC structure. (b-c) $(5 \times 5)$ LEED pattern (b) and schematic drawing (c) of a h-BN layer grown at a temperature of $850°C$ using borazine. (d) LEED pattern of h-BNC grown after heating at $1450 °C$. (e-f) The LEED pattern (e) and schematic drawing (f) of graphene grown after heating at $1600 °C$. All LEED patterns were observed at an electron beam energy of 60 eV.

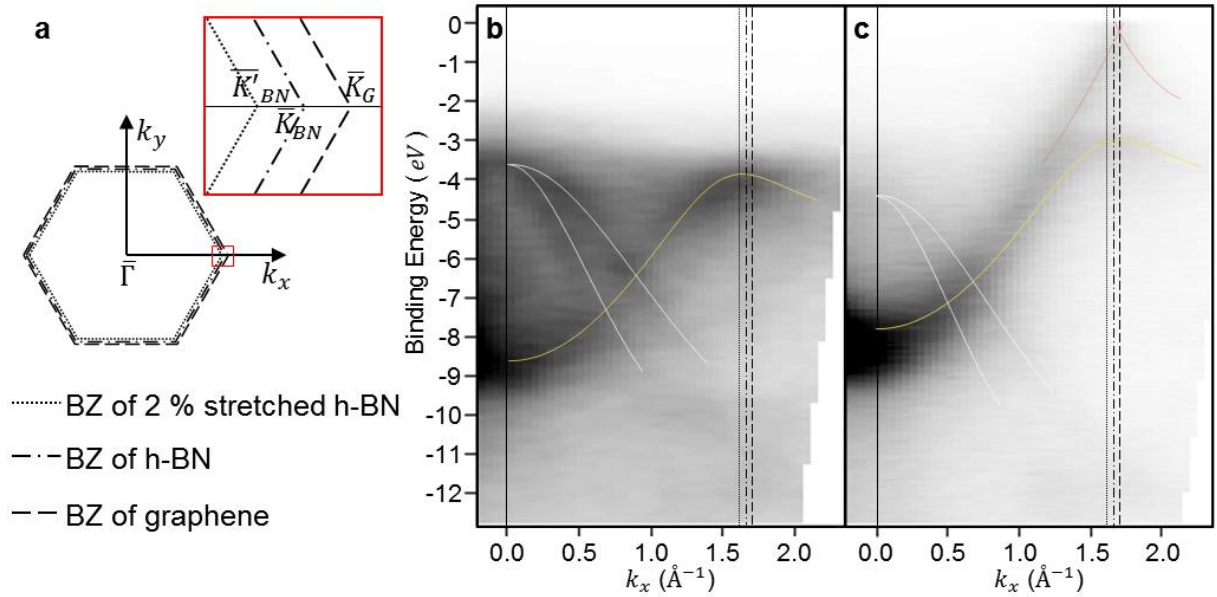

**Figure 2. Electronic band structures of h-BN and h-BNC.** (a) The first Brillion zones of 2% stretched h-BN, unstrained h-BN, and unstrained graphene, denoted by dotted, dash-dot, and dashed lines, respectively. The $\bar{K}$ points of 2% stretched h-BN, unstrained h-BN, and unstrained graphene are denoted by $\bar{K}'_{BN}$, $\bar{K}_{BN}$, and $\bar{K}_G$, respectively. (b,c) Electronic band structures of h-BN (b) and h-BNC (c). The σ and π bands of h-BN in (b) and (c) are indicated by the white and yellow lines, respectively. The π band of graphene in (c) is indicated by the red line.

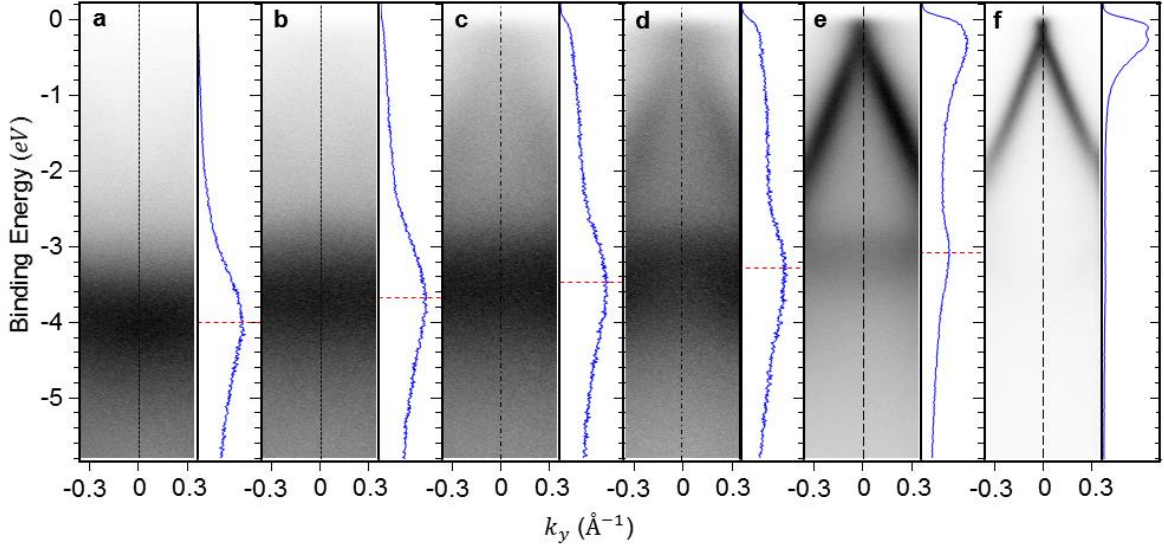

**Figure 3. Changes in electronic band structures along the $k_y$ direction at the K points with increase in temperature.** (a) π band of h-BN on SiC at the $\bar{K}'_{BN}$ point. (b-e) π bands of h-BN and graphene in h-BNC layer at the $\bar{K}'_{BN}$ (b), $\bar{K}_{BN}$ (c, d), and $\bar{K}_G$ (e) points, observed after heating at 1150 °C (a), 1250 °C (b), 1350 °C (c), and 1450 °C (d), respectively. (f) π band of graphene at the $\bar{K}_G$ point. The blue lines are energy distribution curves obtained at $k_y = 0$ Å$^{-1}$ (indicated by black lines in the intensity maps). The red dashed lines indicate the π band maximum of h-BN.

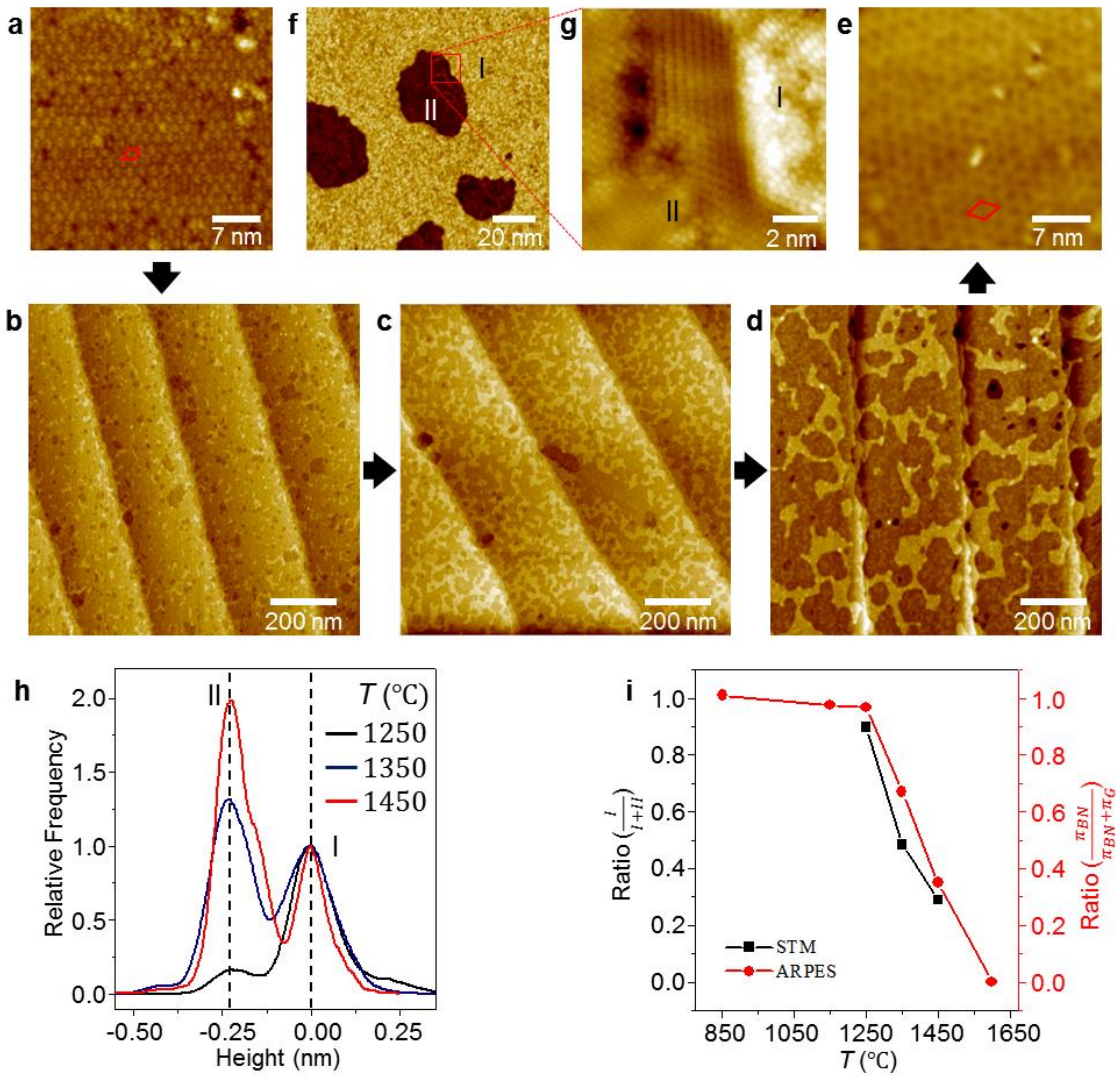

**Figure 4. Filled-state STM images of h-BN, h-BNC and graphene.** (a) STM image of a h-BN layer with a $(5 \times 5)$ superstructure on SiC, where the red lines indicate the $(5 \times 5)$ superstructure. (b-d) STM images of h-BNC acquired after heating at 1250 ℃ (b), 1350 ℃ (c), and 1450 ℃ (d), respectively. (e) STM image of graphene with R0° acquired after heating at 1600 ℃, where the red lines indicate the $(6\sqrt{3} \times 6\sqrt{3})R30°$ superstructure. (f,g) Enlarged STM images of h-BNC acquired after heating at 1250 ℃, where regions I and II in (f) and (g) represent h-BN and graphene, respectively. (h) Relative frequencies plotted against frequency for a 0 nm histogram for h-BNC; the relative frequencies of (b), (c) and (d) are denoted by black, blue and red lines, respectively. (i) The area ratios (black line) of regions I and II in STM images compared with the intensity ratios (red line) of π bands of h-BN and graphene (see Figure 3), as functions of temperature.

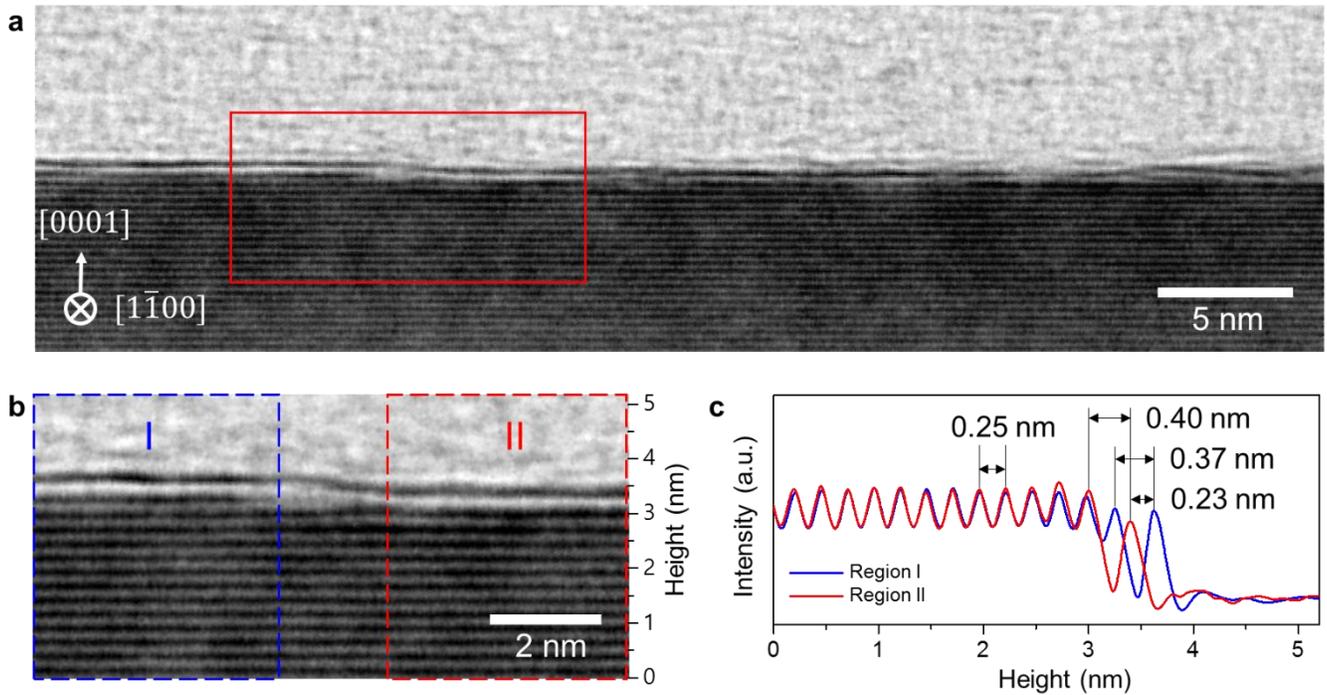

**Figure 5. Cross-section TEM images of h-BNC on SiC(0001).** (a) Large scale cross-section BF-STEM image of h-BNC on SiC(0001) that was prepared after heating at 1250 °C. (b) Enlarged high-resolution BF-STEM image of the solid red rectangle in (a), where h-BN and graphene domains on SiC(0001) are denoted by I and II, respectively. (c) The average line profiles across the layers of the region I (dashed blue rectangle) and II (dashed red rectangle) in (b).

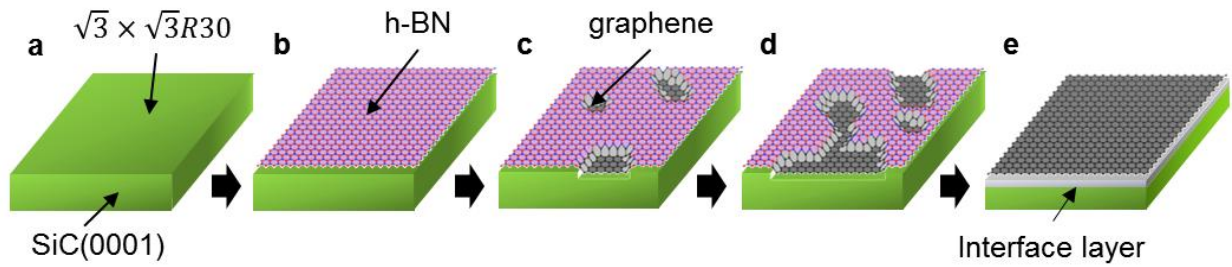

**Figure 6. Schematic drawings of the growth mechanisms of h-BN, h-BNC and graphene.** (a) $(\sqrt{3} \times \sqrt{3})R30°$ superstructure on SiC (0001) (b) h-BN layer grown using borazine at 850 °C. (c-d) h-BNC layer formed after the partial replacement of h-BN with graphene. (e) graphene layer with R0° without h-BN domains grown after heating at 1600 °C. The red, blue, and black spheres denote boron, nitrogen, and carbon atoms, respectively.

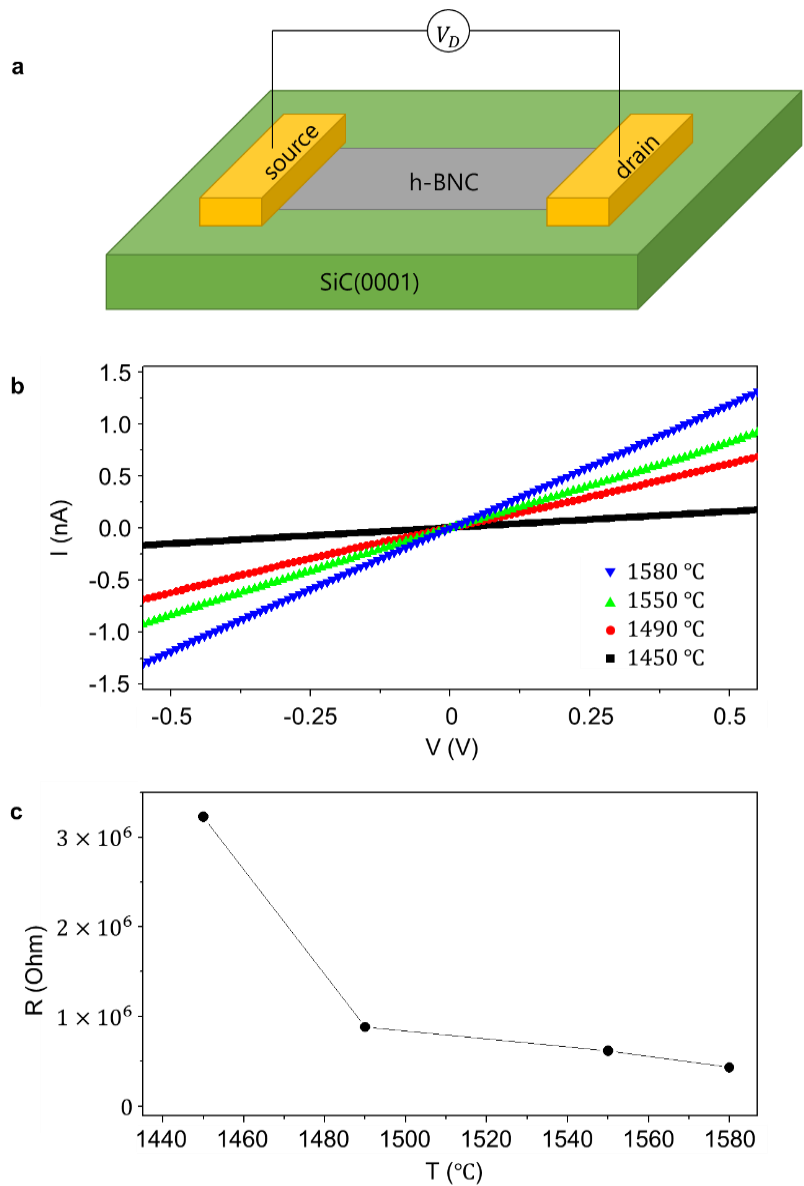

**Figure 7. Resistance measurements of h-BNC devices on SiC(0001)**. (a) A schematic drawing of the h-BNC device. (b) I-V curves of h-BNC devices on SiC(0001), where h-BNC were prepared after heating at 1450, 1490, 1550, and 1580 °C. (c) Resistance determined from the I-V curves in (b).

*Supporting Information*

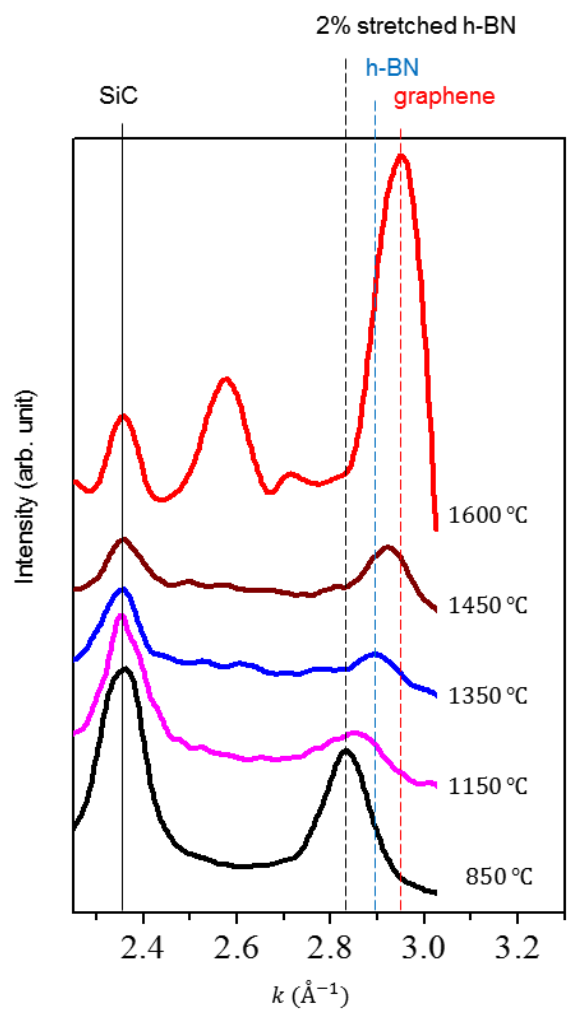

**Figure S1.** Line profiles of LEED patterns along the radial direction after heating to 850 °C, 1150 °C, 1350 °C, 1450 °C and 1600 °C. The peak positions corresponding to bulk SiC, 2% stretched, bare h-BN, and bare graphene are denoted by the solid black line, dashed black, dashed blue, and dashed red line, respectively.

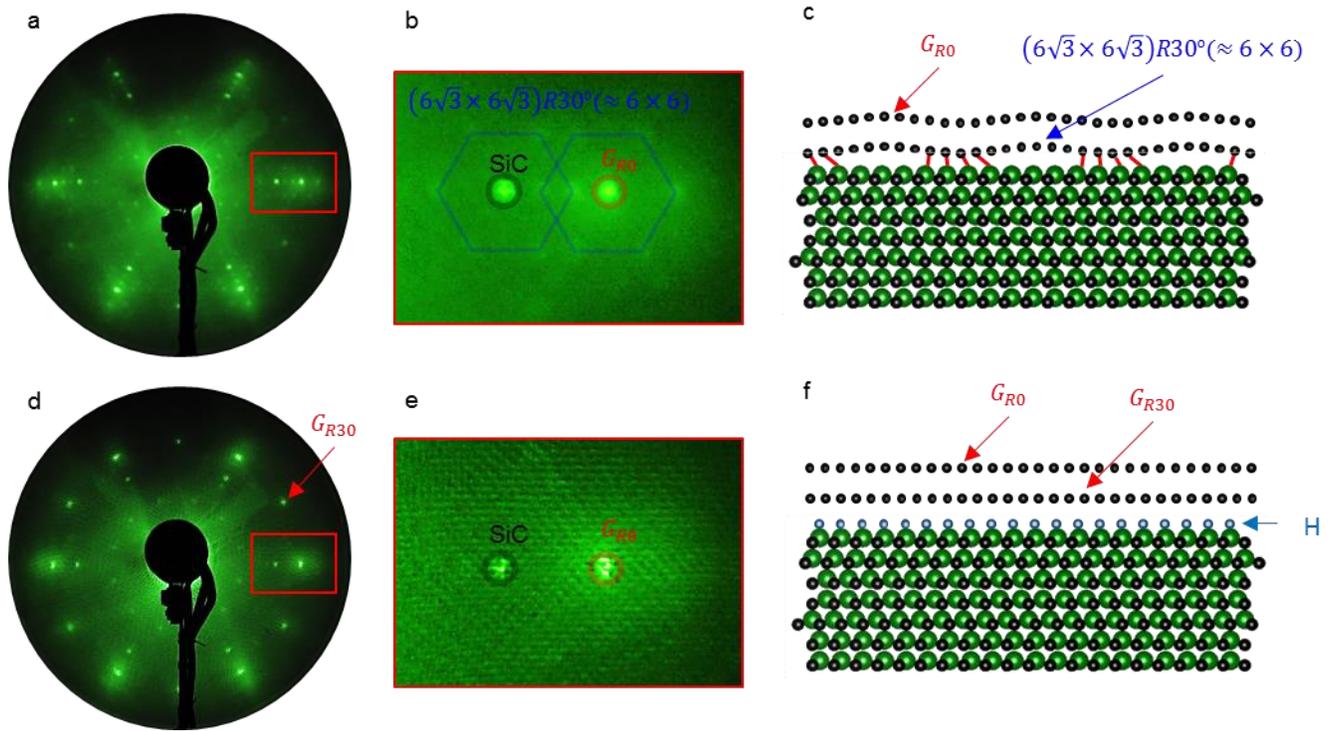

**Figure S2.** (a) and (d) The LEED patterns of graphene with R0° on SiC at an electron energy of 100 eV before and after hydrogen intercalation, respectively. (b) and (e) Enlarged LEED images around bulk SiC and graphene with R0°. (c) and (f) schematic images of graphene structures on SiC with interlayers before and after hydrogen intercalation.

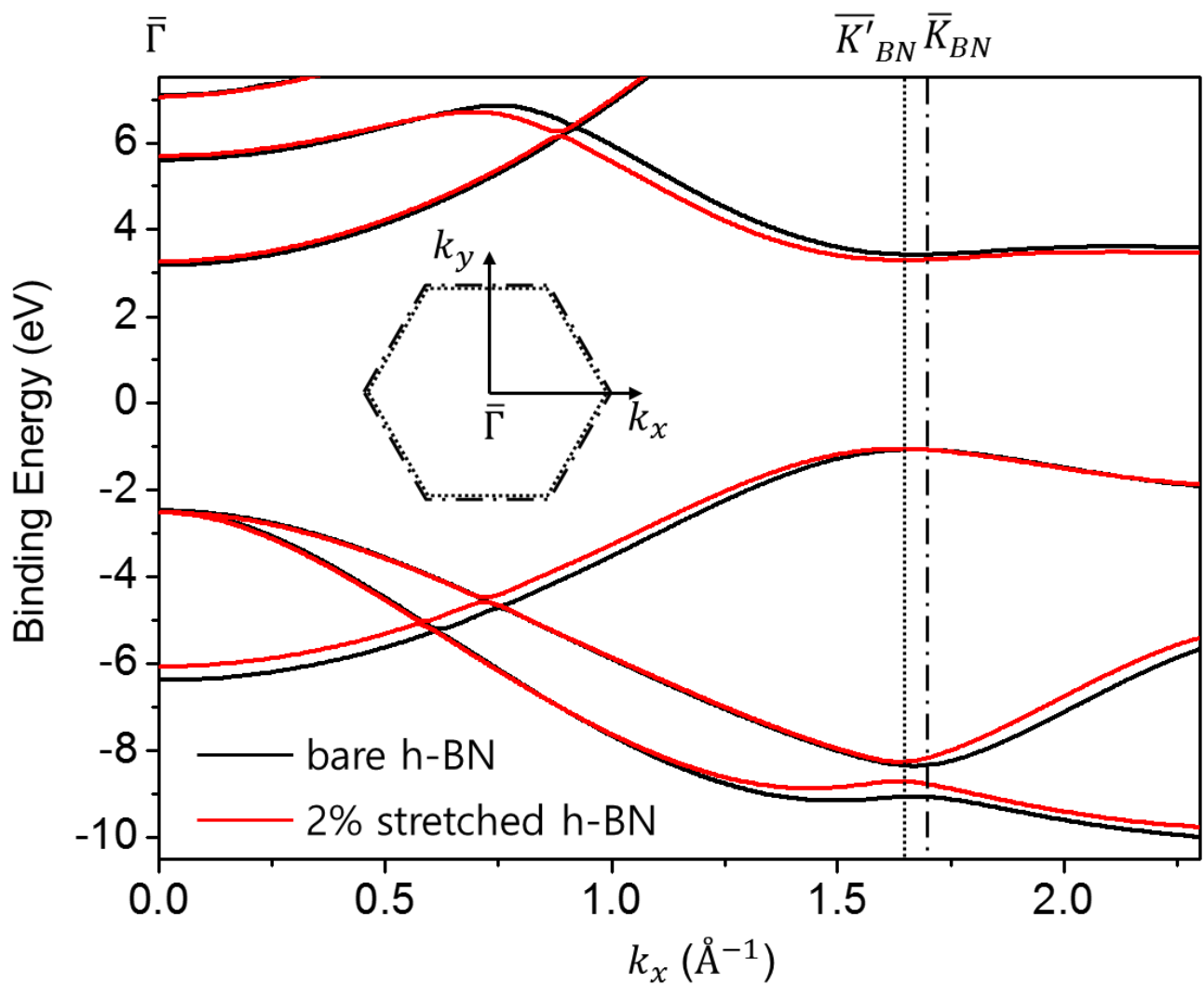

**Figure S3.** Electronic band structures of bare (black lines) and 2%-stretched (red lines) h-BN along the $k_x$ direction. The insert is the first Brillouin zones of bare and 2%-stretched h-BN denoted by dash-dot and dashed lines, respectively. The $\bar{K}$ points of bare and 2%-stretched h-BN are denoted by $\bar{K}_{BN}$ and $\bar{K}'_{BN}$, respectively.

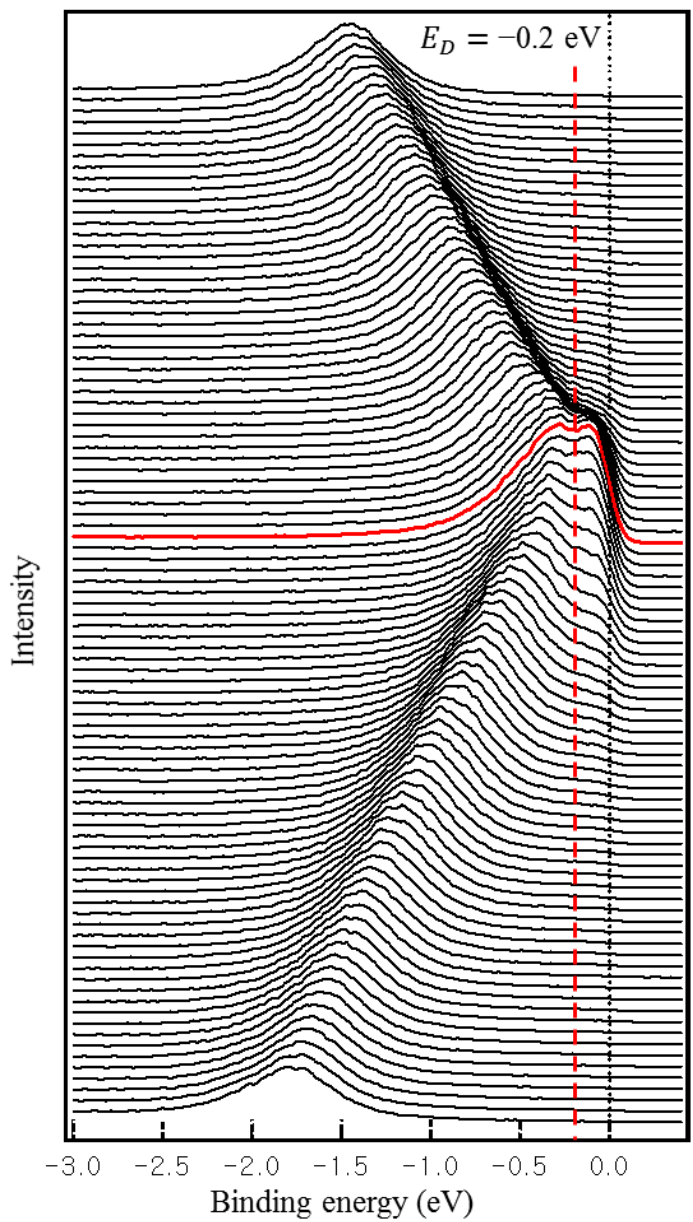

**Figure S4.** EDCs of graphene with R0° along the $k_y$ direction at the $K_G$ point. EDC at the $K_G$ point and the Dirac energy are denoted by red line and red dashed line, respectively.

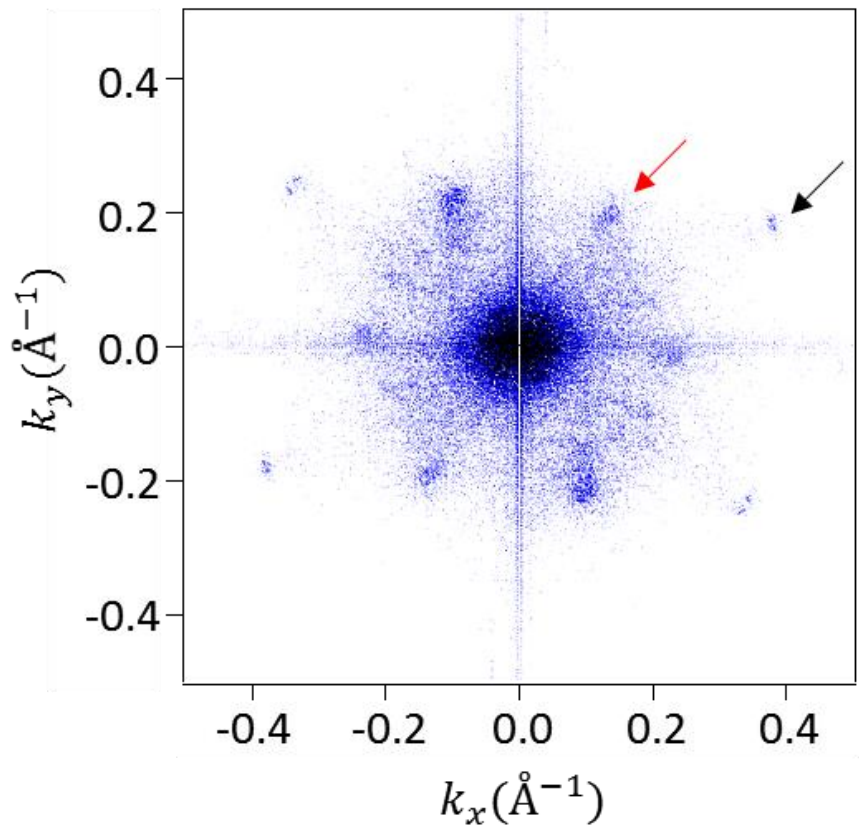

**Figure S5.** A 2D-FFT of a STM image showing $(\sqrt{3} \times \sqrt{3})R30°$ modulation in Figure 4g. The $1 \times 1$ spots of graphene and its $(\sqrt{3} \times \sqrt{3})R30°$ spots induced by the modulation at a boundary are indicated by the black and red arrows, respectively.

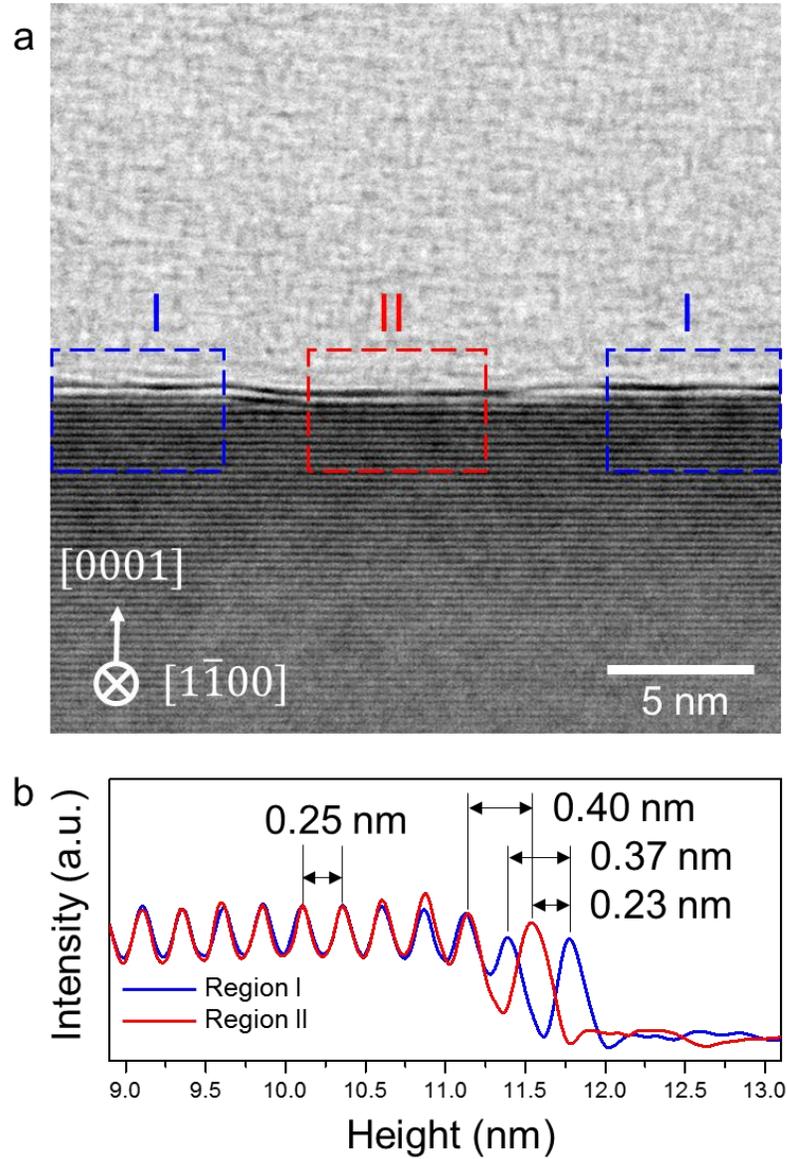

**Figure S6.** A cross-section BF-STEM image of h-BNC. (a) Region I and II denote h-BN and graphene domains on SiC(0001). (b) The average line profiles across the layers of the region I (dashed blue rectangle) and II (dashed red rectangle).

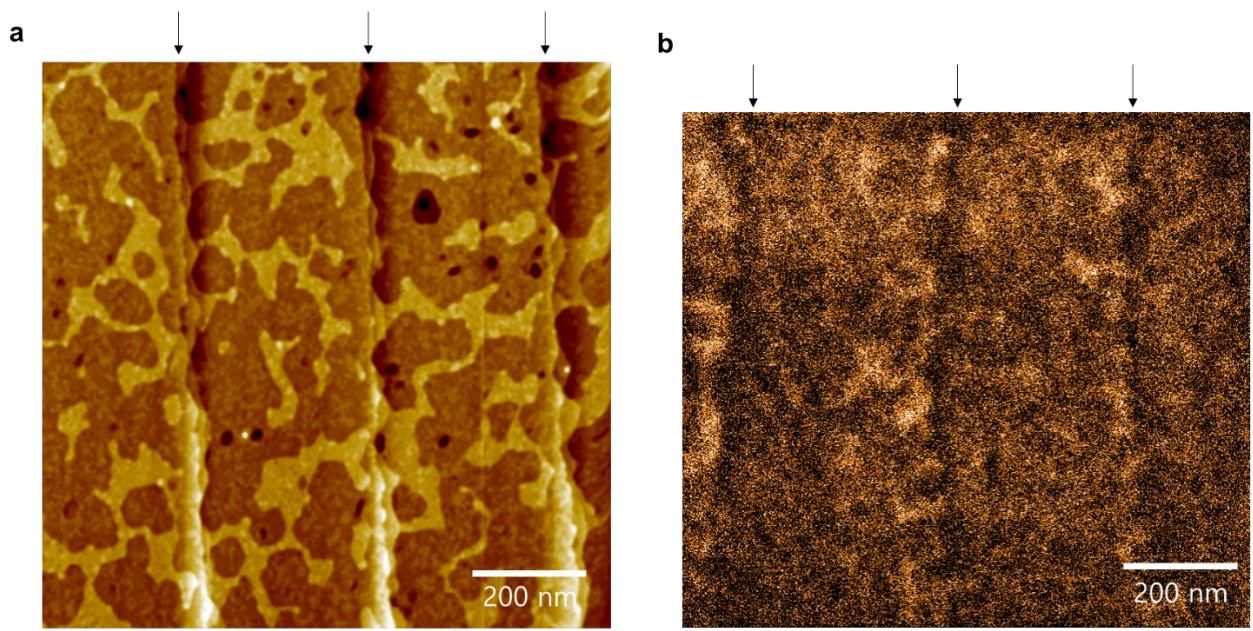

**Figure S7.** Comparison of a STM image with a SEM image. STM (a) and SEM (b) images of h-BNC on SiC(0001) that were prepared after heating at 1450 ℃. Black arrows in (a) and (b) indicate step edges.

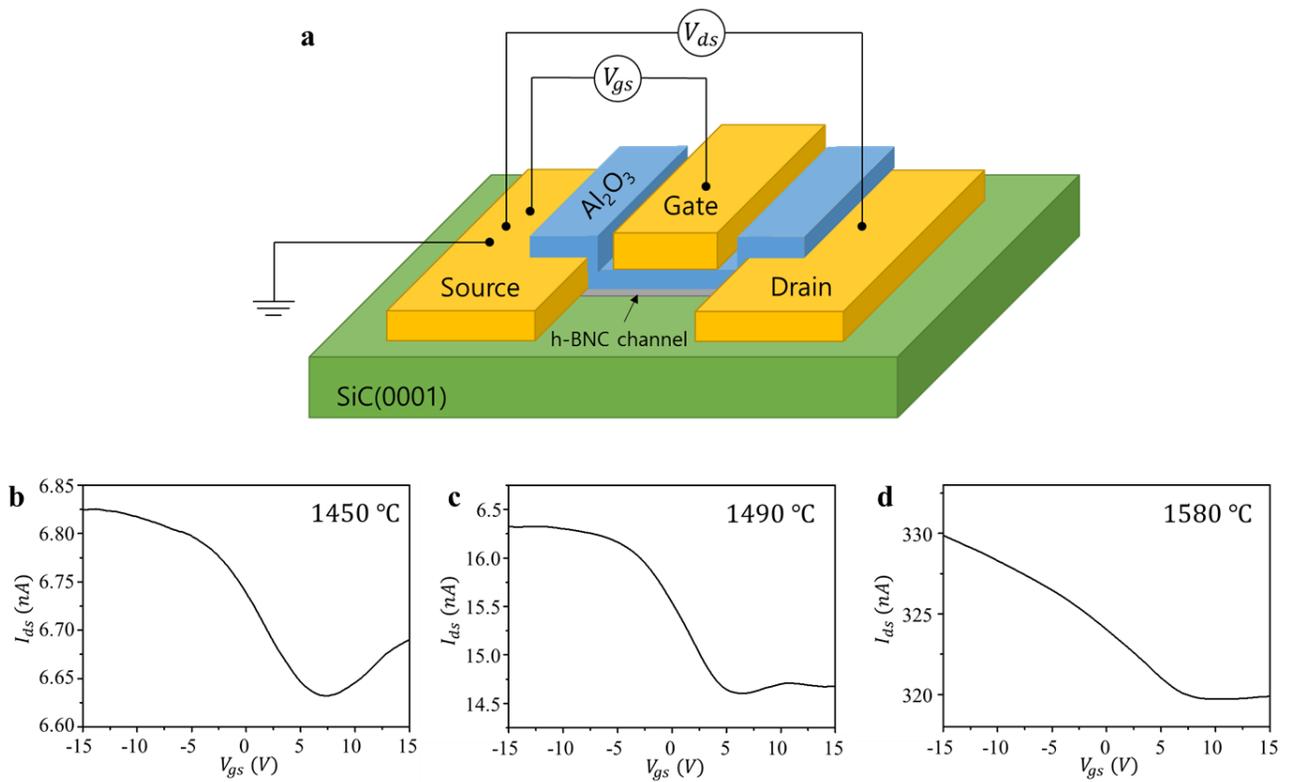

**Figure S8.** (a) A schematic drawing of a h-BNC FET device that was fabricated on a SiC substrate. (b)-(d) The source-drain current as a function of voltage applied to the top gate of the h-BNC FET device after heating at 1450 (b), 1490 (c) and 1580 °C (d). The on/off ratio of the h-BNC device is approximately 1.2 and the mobilities of the FETs are 0.3 (b), 3.1 (c) and 8.9 cm² V⁻¹s⁻¹ (d), respectively.

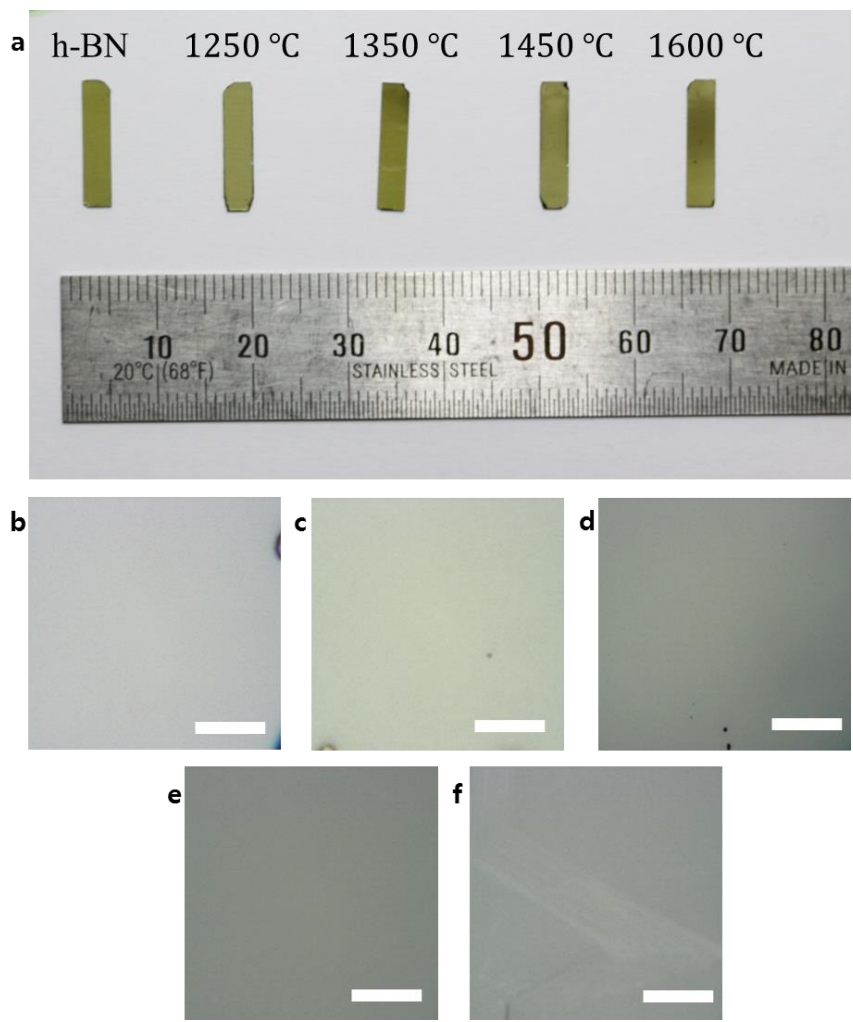

**Figure S9.** (a) Optical images of samples grown on SiC substrates. The enlarged optical images of a h-BN layer (b), h-BNC layers after heating at 1250 (c), 1350 (d), 1450 °C (e) and a graphene layer with R0° after heating at 1600 °C (f). All scale bars are 5 μm.